\documentclass[prb,twocolumn,noshowpacs,amsmath,amssymb]{revtex4}
\usepackage{graphicx}
\usepackage{dcolumn}
\usepackage{bm}
\usepackage{amsmath}
\usepackage{amsxtra}
\usepackage{tabularx}
\def\beq#1\eeq{\begin{equation}#1\end{equation}}
\def\beqry#1\eeqry{\begin{eqnarray}#1\end{eqnarray}}

\newcommand{\sml}[1]{ \scriptscriptstyle #1 }

\newcommand{\bsigma}{\mbox{\boldmath $\sigma$  \unboldmath} }

\newcommand{\bnabla}{\mbox{\boldmath$\nabla$\unboldmath}}

\newcommand{\bu}{{\bf u}}
\newfont{\ccc}{cmbxsl10 at 10pt}

\begin{document}

\title{
% Noncollinear Spins in Density Functional Calculations.\\
Noncollinear Magnetism in Density Functional Calculations}
\author{Juan E. Peralta}
\author{Gustavo E. Scuseria}
\affiliation{Department of Chemistry, Rice University, Houston, Texas 77005}
\author{Michael J. Frisch}
\affiliation{Gaussian Inc., 340 Quinnipiac St., Bldg. 40, Wallingford, Connecticut 06492}

\date{\today}

%%%%%%%%%%%%%%%%%%%%%%%%%%%%%%%%%%%%%%%%%%%%%%%%%%%%%%%%%%%%%%%%%%%%%%%%%%%%%%%%%%%%%%%%%%%%%%
\begin{abstract}
We generalize the treatment of the electronic spin degrees 
of freedom in density functional calculations to
the case where the spin vector variables employed in the definition of the energy functional 
can vary in any direction in space.
The expression for  the generalized exchange-correlation potential matrix elements is derived for general functionals
which among their ingredients include the electron density, its gradient and Laplacian, the kinetic energy density, 
and non-local Hartree-Fock type exchange. 
We present calculations on planar Cr clusters that exhibit ground states with noncollinear spin densities 
due to geometrically frustrated  antiferromagnetic interactions.
\end{abstract}
%%%%%%%%%%%%%%%%%%%%%%%%%%%%%%%%%%%%%%%%%%%%%%%%%%%%%%%%%%%%%%%%%%%%%%%%%%%%%%%%%%%%%%%%%%%%%%


\maketitle

%%%%%%%%%%%%%%%%%%%%%%%%%%%%%%%%%%%%%%%%%%%%%%%%%%%%%%%%%%%%%%%%%%%%%%%%%%%%%%%%%%%%%%%%%%%%%%
\section{Introduction}
%%%%%%%%%%%%%%%%%%%%%%%%%%%%%%%%%%%%%%%%%%%%%%%%%%%%%%%%%%%%%%%%%%%%%%%%%%%%%%%%%%%%%%%%%%%%%%

Since the spin polarized formulation\cite{Barth_1972} of density functional theory (DFT)\cite{kohn64,kohn65} 
was introduced, a large
number of applications have been carried out in magnetic systems. In most cases, the spin density 
is assumed to adopt a single direction  (collinear) at each point in space  that is usually taken as $z$. However, there are
a number of systems where the spin density (or magnetization density) can take a more complicated 
structure, and vary its {\em direction} at each 
point in space. These type of noncollinear structures were observed in 
the form of helical spin density waves or spin spirals for the ground state of $\gamma$-Fe,\cite{Tsunoda_1989,Sjostedt_2002}
in geometrically frustrated systems
like for instance the Kagom\'e antiferromagnetic lattice\cite{Grohol_2005}, and in systems with competing magnetic 
interactions such as the composite magnet LaMn$_2$Ge$_2$\cite{Venturini_1994} and Fe$_{\sml 0.5}$Co$_{\sml 0.5}$Si.\cite{Uchida_2006}

Several papers dealing with noncollinear spin density in DFT calculations have been published in the literature. 
The pioneer work of K\"ubler and coworkers\cite{Kubler_1988} for the noncollinear 
local spin density approximation (LSDA) was later followed by a number of independent implementations 
and applications. Most of these
implementations were carried out using periodic boundary conditions and plane waves,
and are based either  on  the LSDA\cite{Nordstrom_1996,Oda_1998} or on  a  generalized gradient approximation  (GGA).\cite{Kurz_2004}
Yamanaka and coworkers developed a generalized DFT code based on Gaussian type orbitals.\cite{Yamanaka_2000}
Some noncollinear DFT calculations have been published dealing with magnetic crystals,\cite{Nordstrom_1996,Sjostedt_2002}
and with fourth period transition metal clusters.\cite{Oda_1998,Kohl_1999,Longo_2005,Mejia_2006} 
%Although not formally justified, 
In all cases, the realization of the LSDA and GGA employed in noncollinear 
calculations is  the same as that developed for collinear spin systems.  

Different parametrizations of the exchange-correlation energy ($E_{xc}$) have been proposed beyond the LSDA and the GGA, 
incorporating more ingredients in the definition of $E_{xc}$.
The third rung in this hierarchy\cite{JacobsLadder} is the meta-GGA, which includes the kinetic energy density 
as a functional ingredient. Also, hybrid density functionals, which contain a portion of
Hartree-Fock type exchange, can be regarded as belonging to the fourth rung (hyper-GGA) in this picture.\cite{JacobsLadder}
The purpose of this paper is
to provide a consistent generalization for the treatment of noncollinear spin variables in DFT calculations 
beyond the LSDA.

%%%%%%%%%%%%%%%%%%%%%%%%%%%%%%%%%%%%%%%%%%%%%%%%%%%%%%%%%%%%%%%%%%%%%%%%%%%%%%%%%%%%%%%%%%%%%%
\section{Theory}
%%%%%%%%%%%%%%%%%%%%%%%%%%%%%%%%%%%%%%%%%%%%%%%%%%%%%%%%%%%%%%%%%%%%%%%%%%%%%%%%%%%%%%%%%%%%%%

To allow for noncollinear spin states in density functional calculations, we start by introducing 
two-component spinors as Kohn-Sham (KS) orbitals:
\begin{eqnarray}
% \Psi_i = \psi_i^{\alpha}\alpha + \psi_i^{\beta}\beta\,,
\Psi_i = \left(
\begin{array}{c}
\psi_i^{\alpha}\\
\psi_i^{\beta}
\end{array}
\right)
\label{Eq:Psi}
\,,
\end{eqnarray}
where $\psi_i^\alpha$ and $\psi_i^\beta$ are spatial orbitals that 
can be expanded in a linear combination of atomic orbitals,
\begin{equation}
\psi_i^\sigma({\bf r}) = \sum_\mu c_{\mu i }^{\sigma} \phi_\mu
({\bf r})~~~~~~~~~~~ (\sigma = \alpha, \beta)
\label{Eq:Spinors}
\,.
\end{equation}
% Note that the two-component spinors $\Psi_i $ are not necessarily eigenfunctions of any component of 
% the spin operator.

Using the KS formulation, the electronic energy is partitioned into four contributions:
\begin{equation}
E= E_T + E_N + E_J + E_{xc}\,,
\end{equation}
where $E_T$ is the kinetic energy, $E_N$ is the nuclear-electron interaction energy,
$E_J$ is the classical electron-electron Coulomb repulsion energy, and $E_{xc}$ is the exchange-correlation (XC) energy.
Searching for stationary solutions of $E$ is equivalent to solving the  KS equations, which in 
terms of two-component spinors $\Psi_i$ read:
\begin{equation}
(T + V_N + J + V_{xc})\Psi_i = \epsilon_i \Psi_i\,,
\label{eq:KS}
\end{equation}
where $T=-1/2\,\nabla^2$ is the kinetic energy operator, $V_N$ is the external electron-nuclear potential, 
$J$ is the Coulomb operator, and $V_{xc}$ is the exchange-correlation (XC) potential.
We shall here refer to the KS equations in a two-component spinor basis as the generalized KS (GKS) equations. 
Since $T$, $V_N$, and $J$ are diagonal in the two-dimensional spin space, the only 
term in Eq.~\ref{eq:KS} that couples $\psi_i^{\alpha}$ and  $\psi_i^{\beta}$  is $V_{xc}$ (the spin-orbit operator,
present in a relativistic  Hamiltonian, also couples the two spinor components).
The potential $V_{xc}$ depends on the choice of $E_{xc}$ and therefore the coupling between 
 $\psi_i^{\alpha}$ and  $\psi_i^{\beta}$ in the non-relativistic GKS equations depends exclusively on $E_{xc}$.

Let us first recall the standard formulation of $E_{xc}$ commonly employed in (collinear) unrestricted 
KS (UKS) calculations.
The general expression of $E_{xc}$ for a hybrid case can be cast as:\cite{Becke1993}
\begin{equation}
E_{xc} = a  E^{\sml DFA}_x + E^{\sml DFA}_c + (1-a) E^{\sml HF}_x  \,,
\label{eq:Exc}
\end{equation}
where $E^{\sml DFA}_{x}$ and $E^{\sml DFA}_{c}$ are the exchange-correlation contributions to the energy 
at some (semi)local density functional approximation (DFA), respectively, 
$E^{\sml HF}_x$ is the Hartree-Fock type exchange energy, and $a$ is a mixing parameter ($0\le a \le1$). The functional
forms of $E^{\sml DFA}_x$ and $E^{\sml DFA}_c$ (as well as the parameter $a$) depend, of course, on the choice of
the functional employed in the actual calculation. A general expression for $E^{\sml DFA}_{xc}=a E^{\sml DFA}_x + E^{\sml DFA}_c$ can be 
written as
\begin{eqnarray}
E_{xc}^{\sml DFA}=
\int d^3r\,
f({ \cal Q }) \label{exc} \,,
\end{eqnarray}
where ${\cal Q} $ is a set of  variables (included in the definition of $E_{xc}$):
\begin{eqnarray}
{\cal Q} \equiv 
\Big\{ n_{\alpha},
\, n_{\beta },
\, {\bnabla} n_{\alpha},
\, {\bnabla} n_{\beta},
\, \tau_{\alpha},
\, \tau_{\beta}, 
\, \nabla^2 n_{\alpha} 
\, \nabla^2 n_{\beta} 
\Big\} \,.
\end{eqnarray}
Here $n_\alpha$ and $n_\beta$ are the $\alpha$ and $\beta$ electron densities, and $\tau_{\alpha}$ and $\tau_{\beta}$ are the
$\alpha$ and $\beta$ kinetic energy densities, respectively, representing the ``up'' and ``down'' 
components along the $z$ axis. 

On the other hand, in the noncollinear case, where the vector component of the local  variables employed in the definition of 
$E_{xc}$ can point in any direction,
$E_{xc}^{\sml DFA}$  can be generalized as follows,
\begin{eqnarray}
E^{\sml NC}_{xc}=
\int d^3r\,
f({ \cal \widetilde{Q}}) \label{Exc_nc} \,,
\end{eqnarray}
where 
\begin{eqnarray}
{\cal \widetilde{Q} } \equiv 
\Big\{ n_{\sml +},
\, n_{\sml - },
\,  {\bnabla} n_{\sml +},
\,  {\bnabla} n_{\sml -},
\, \tau_{\sml + },
\, \tau_{\sml - }, 
\, \nabla^2 n_{\sml +}
\, \nabla^2 n_{\sml -}
\Big\} \,.
\end{eqnarray}
Here the subindices $+$ and $-$ refer to variables expressed in a local reference frame along the 
local spin quantization axis. The definition of these variables is given in detail in Section II.
Note that by replacing $\cal Q$ by $\cal \widetilde{Q}$, we have only added degrees 
of freedom to the local variables in such a way that they are compatible with
any arbitrary choice of the local spin axis. The dependence of $f$ (and therefore $E^{\sml NC}_{xc}$) 
on these variables remains unchanged. 

Two conditions must be satisfied by the set of variables in $\cal \widetilde{Q}$.
First, we  should recover the standard collinear case if we allow spin polarization only in one direction, and therefore
replace the labels $+$ and $-$  in $E^{\sml NC}_{xc}$ by $\alpha$ and $\beta$, respectively. In other words, 
noncollinear GKS solutions should coincide with collinear solutions obtained with the standard UKS approximation 
in cases where the ground state solution is collinear. Second,
for spin-independent Hamiltonians, any arbitrary choice (other than $z$) 
of the spin quantization axis should leave the energy unchanged. This means that
any rigid rotation of all local reference frames should not change the total energy.

% The Hartree-Fock type exchange contribution to $E_{xc}$, needed for hybrid density functional calculations, 
% can be easily generalized in terms of 
% two-component spinors [for instance, see the generalized unrestricted HF (GUHF) theory in Ref.~\onlinecite{McWeeny}]. 

We therefore assume that  $E^{\sml NC}_{xc}$  depends on the local variables $+$ and $-$ in the same manner as in
the standard collinear case, and that  $E^{\sml NC}_{xc}$   must be invariant under rigid rotations of
the spin quantization axis. This is, of course, not the most general possibility to define energy functionals for
noncollinear magnetic systems.\cite{Kleinman_1999}

For practical applications, it is necessary to evaluate the XC potential matrix
to be employed in the solution of the GKS equations. These matrix elements 
in a set of localized orbitals $\{\phi_\xi\}$  can be written as:
\begin{eqnarray}
(V_{xc}^{\sml NC})_{\mu\nu}& =  &  \nonumber
\int d^3r\, \frac{\partial f (\cal \widetilde{Q})}{\partial P_{\mu\nu}} \\ 
& = &  \sum_{p, q \in \cal \widetilde{Q}} \int d^3r\, \frac{\partial f (\cal \widetilde{Q})}{\partial q} 
\frac{\partial q}{\partial p } 
\frac{\partial p}{\partial P_{\mu\nu}} \,,
\label{Eq:Potential}
\end{eqnarray}
where $P_{\mu\nu}$ are matrix elements of the generalized density matrix,
\begin{eqnarray}
P_{\mu\nu} = \sum_{i \in  occ} 
 \left( \begin{array}{cc}
c_{\mu i }^\alpha c_{\nu i }^{\alpha {\ast}}  & c_{\mu i }^\alpha c_{\nu i }^{\beta {\ast}}  \\
c_{\mu i }^\beta c_{\nu i }^{\alpha {\ast}}  &   c_{\mu i }^\beta c_{\nu i }^{\beta {\ast}}   \end{array} \right)\,,
\label{Eq:GP}
\end{eqnarray}
and $p$ represents variables that are {\em linear} in $P_{\mu\nu}$. The derivative $\partial f / \partial q $ 
remains the same as in the collinear case. 
The rest of this section is devoted to the definition of the variables $q \in \cal \widetilde{Q}$ in the local reference frame,
and to the evaluation of $(V_{xc}^{\sml NC})_{\mu\nu}$.

%%%%%%%%%%%%%%%%%%%%%%%%%%%%%%%%%%%%%%%%%%%%%%%%%%%%%%%%%%%%%%%%%%%%%%%%%%%%%%%%%%%%%%%%%%%%%%
\subsection{Density}
\label{SSec:LSDA}
%%%%%%%%%%%%%%%%%%%%%%%%%%%%%%%%%%%%%%%%%%%%%%%%%%%%%%%%%%%%%%%%%%%%%%%%%%%%%%%%%%%%%%%%%%%%%%

The main ingredient for the construction of $ E_{xc}^{\sml NC} $  in the LSDA is the generalized density, $\bar n$, which
can be written in a two-component spin space  as
\begin{eqnarray}
{\bar n } = \frac{1}{2}( {n} + {\bf m}\cdot \bsigma )
 = \frac{1}{2} \left( \begin{array}{cc} 
 n +  m_z & m_x - i m_y \\
 m_x + i m_y  & n - m_z \end{array} \right)  \,,
\label{Eq:GN}
\end{eqnarray}
where $n$ is the electron density, ${\bf m}=(m_x, m_y,m_z)$ is the spin density (magnetization) vector, 
and $\bsigma=(\sigma_x, \sigma_y, \sigma_z)$ are the Pauli matrices.
In terms of two-component spinors $\Psi_i$, $n$ and ${\bf m}$ are defined as
\begin{eqnarray}
n({\bf r}) = \sum_{i \in occ} \Psi_i^\dagger ({\bf r}) 
\Psi_i({\bf r})\,,
\end{eqnarray}
and
\begin{eqnarray}
m_k({\bf r}) = 
\sum_{i \in occ} \Psi_i^\dagger ({\bf r}) \sigma_k
\Psi_i({\bf r})\,\,\,\,\,\,\,\,\,\,\,(k=x,y,z).
\end{eqnarray}
Using Eqs.~\ref{Eq:Psi}, \ref{Eq:Spinors}, and \ref{Eq:GP}, it is straightforward to 
express  $n$ and ${\bf m}$ as a linear combination of matrix elements of the generalized density matrix $P_{\mu\nu}$.
%  where the linear dependence of $n$ and ${\bf m}$ with the generalized density matrix is evident.

A local reference system where the generalized density ${\bar n}$ is diagonal can be 
obtained by rotating ${\bar n}$ into $\bar n \,'$,
\begin{eqnarray}
{\bar n \,'}
 = \left( \begin{array}{cc}  
 n{\sml +} & 0 \\
 0 & n{\sml -}  \end{array} \right)  \,,
\end{eqnarray}
where 
\begin{eqnarray}
 n_{\sml \pm} = \frac{1}{2}(n \pm m) =  \frac{1}{2}\big(\, n \pm \sqrt{m_x^2 + m_y^2 + m_z^2}\, \big) 
\end{eqnarray}
are the eigenvalues of $\bar n$.
The densities $ n_{\sml +} $ and $ n_{\sml -}$  can be regarded as the local
analogous of $ n_{\alpha}$ and $ n_{\beta} $. 
The potential  $ V_{xc}^{\sml NC} $ is
\begin{eqnarray}
V_{xc}^{\sml NC} & =  & \frac{\delta E^{\sml NC}_{xc} }{\delta \bar n} \nonumber \\
& = & \frac{1}{2} (f^+ + f^- {\hat{\bf m}}\cdot{\bsigma} ) \,,
\label{Eq:VLSDA}
\end{eqnarray}
where
\begin{eqnarray}
f^\pm=\frac{\partial f}{ \partial n_{\sml{+}}}  \pm
\frac{\partial f}{ \partial n_{\sml{-}}} \,,
\end{eqnarray}
and ${\hat{\bf m}}={\bf m}/m$ is the unit vector in the direction of ${\bf m}$. The matrix elements of  $V_{xc}^{\sml NC}$ can
be evaluated straightforwardly as
\begin{eqnarray}
(V_{xc}^{\sml NC})_{\mu\nu}=
\int d^3r\,\phi_\mu V_{xc}^{\sml NC} \phi_\nu\,.
\end{eqnarray}

The potential $V_{xc}^{\sml NC}$ 
in Eq.~\ref{Eq:VLSDA} can be split into two contributions,
\begin{eqnarray}
V_{xc}^{\sml NC} = {\cal E}_{xc}^{\sml NC} + {\bsigma} \cdot {\cal {\bm B}}_{xc}^{\sml NC}\,, 
\end{eqnarray}
where $  {\cal E}_{xc}^{\sml NC}=f^+/2$ can be interpreted as a scalar (electrostatic) potential 
and ${\cal{\bm B}}_{xc}^{\sml NC}=f^- \hat{\bf m}/2$ as a spin-dependent (magnetic) potential,
which for the LSDA is always parallel to ${\bf m}$. It is worth mentioning that in the limit 
of no magnetization ($m\rightarrow0$), $f^-\rightarrow0$ and therefore ${\cal{\bm B}}_{xc}^{\sml NC}\rightarrow {\bf 0}$,
recovering the non-magnetic case.

%%%%%%%%%%%%%%%%%%%%%%%%%%%%%%%%%%%%%%%%%%%%%%%%%%%%%%%%%%%%%%%%%%%%%%%%%%%%%%%%%%%%%%%%%%%%%%
\subsection{Density gradients}
\label{SSec:GGA}
%%%%%%%%%%%%%%%%%%%%%%%%%%%%%%%%%%%%%%%%%%%%%%%%%%%%%%%%%%%%%%%%%%%%%%%%%%%%%%%%%%%%%%%%%%%%%%

Let us consider next GGA energy functionals.
In this case the new ingredients in $E_{xc}^{\sml NC}$ are $\bnabla n_+$ and $\bnabla n_-$, 
whose Cartesian component $j$ ($j=x, y, z$) is
\begin{eqnarray}
\nabla_j n_{\pm} & = & \frac{\partial n_{\pm}}{\partial j} \nonumber  \\
& = & \frac{1}{2}\Big( \nabla_j n \,  \pm \,  \frac{1}{m}\sum_{k=x,y,z} m_k \nabla_j m_k   \Big)
\,. \label{Eq:grho}
\end{eqnarray}
We note in passing that this family of density functionals usually depends on the gradient of the density
through the auxiliary quantities\cite{Pople1992}
\begin{eqnarray}
\gamma_{ab}=\bnabla n_a \cdot \bnabla n_b \,\,\,\,\,\,\,\,\, a,b=+,- \,.
\end{eqnarray}

%To obtain the expression of the matrix elements of $V_{XC}^{\sml{NC}}$,
%three contributions to the derivative of $f$ with respect to the variables $\bnabla n$, $\bf m$, and  $\bnabla {\bf m}$
%must be taken into account,
The matrix elements of $V_{xc}^{\sml{NC}}$  are 
\begin{eqnarray}
(V_{xc}^{\sml{NC}})_{\mu\nu} = & &
\sum_{j=x,y,z} \int \frac{\partial f}{\partial (\nabla_j n)}\nabla_j(\phi_\mu \phi_\nu)d^3r \nonumber \\
& & + \sum_{j,k=x,y,z} \int \frac{\partial f }{\partial (\nabla_j m_k )} \sigma_k   \nabla_j(\phi_\mu \phi_\nu)d^3r \nonumber \\
 & & + \sum_{k=x,y,z} \int  \frac{\partial f }{\partial m_k} \sigma_k (\phi_\mu \phi_\nu) d^3r
\,, \label{Eq:VGGA}
\end{eqnarray}
where the first term on the right-hand side of Eq.~\ref{Eq:VGGA} contributes to the scalar potential 
${\cal E}_{xc}^{\sml NC}$, and the second and last terms add to the magnetic potential ${\cal{\bm B}}_{xc}^{\sml NC}$.
The last term on the right-hand side of Eq.~\ref{Eq:VGGA} arises from the 
fact that $\bnabla n_{\pm}$ depends on ${\bf m}$ (Eq.~\ref{Eq:grho}).
Applying the chain rule, and making use of Eq.~\ref{Eq:grho},
the derivatives of $f$ % through $\bnabla n_{\pm}$ 
in Eq.~\ref{Eq:VGGA} can be expressed as (we do not consider here derivatives of  $f$ arising from $\partial f/\partial n_{\pm}$ 
since they where considered in the previous section)
\begin{eqnarray}
\frac{\partial f }{\partial (\nabla_j n)} =
\frac{1}{2}
g_j ^{\sml +}\,,\label{Eq:GGA1}
\end{eqnarray}
\begin{eqnarray}
\frac{\partial f }{\partial (\nabla_j m_k)} =
\frac{1}{2} \frac{m_k}{m}
g_j^{\sml -} \,,\label{Eq:GGA2}
\end{eqnarray}
and
\begin{eqnarray}
\frac{\partial f }{\partial m_k} =
\sum_{l=x,y,z} \frac{1}{2} \Big\{ 
\frac{\nabla_l m_k}{m} 
g_l^{\sml +} - 
\frac{m_k}{m^2} g_l^{\sml -}  \nabla_l m
\Big\}\,,\label{Eq:GGA3}
\end{eqnarray}
where we have defined
\begin{eqnarray}
g_k^{\pm} = 
\frac{\partial f }
{\partial (\nabla_k n_{\sml +})}  \pm 
                  \frac{\partial f }
{\partial (\nabla_k n_{\sml -})}
\,.
\end{eqnarray}

Alternatively, one can calculate $V_{xc}^{\sml{NC}}$  as the functional derivative
\begin{eqnarray}
V_{xc}^{\sml NC} & =  & \frac{\delta E^{\sml NC}_{xc} }{\delta \bar n} \nonumber \\
& = & \frac{\delta E^{\sml NC}_{xc} }{\delta n}  + \sum_i \frac{\delta E^{\sml NC}_{xc} }{\delta m_i} \sigma_i. \label{fdergga}
\end{eqnarray}
Using the chain rule for functional derivatives, $\delta E^{\sml NC}_{xc}/\delta m_i$ can be written as
\begin{eqnarray}
\frac{\delta E^{\sml NC}_{xc} }{\delta m_i} = \frac{1}{2}(\frac{\delta E^{\sml NC}_{xc} }{\delta n_+} - 
\frac{\delta E^{\sml NC}_{xc} }{\delta n_-} ) \frac{m_i}{m}\,. \label{chain}
\end{eqnarray}
Multiplying Eq.~\ref{chain} by $ \phi_\mu \phi_\nu $, integrating over all space and applying integration by parts, 
we obtain Eq.~\ref{Eq:VGGA}. 
Therefore, from our definition of  $E^{\sml NC}_{xc}$, the contribution to the XC magnetic field
$ {\cal{\bm B}}_{xc}^{\sml NC} $ from Eq.~\ref{fdergga} is always parallel to ${\bf m}$. However,
this is not necessarily the case for a general form of a GGA functional, as 
% The properties of the XC spin torque were 
shown by Capelle and coworkers.\cite{capelle} 

One issue that is worth addressing is how our formulation differs from previous noncollinear generalizations of the GGA.
In our case, we employ $\bnabla n_{\sml \pm}$ as defined in Eq.~\ref{Eq:grho}, and therefore the ingredients
used in  $E_{xc}^{\sml{NC}}$ are strictly the gradients of the quantities $n_{\sml \pm}$. 
Other implementations\cite{Knopfle_2000,Kurz_2004} employ either $\bnabla m$ or the $z$ component of the
projection of $\bnabla m_k$ onto $\bf m$, therefore imposing the constraint of 
$ {\cal{\bm B}}_{xc}^{\sml NC} $ being parallel to $\bf m$. 
Here as in the LSDA case, in the limit
of no magnetization ($m\rightarrow0$ and $\bnabla m\rightarrow {\bf 0}$), $g_k^-\rightarrow0$ for $k=x,y,z$ 
and hence ${\cal{\bm B}}_{xc}^{\sml NC}\rightarrow {\bf 0}$,
recovering the non-magnetic case.

%%%%%%%%%%%%%%%%%%%%%%%%%%%%%%%%%%%%%%%%%%%%%%%%%%%%%%%%%%%%%%%%%%%%%%%%%%%%%%%%%%%%%%%%%%%%%%
\subsection{Kinetic energy density}
\label{SSec:TAU}
%%%%%%%%%%%%%%%%%%%%%%%%%%%%%%%%%%%%%%%%%%%%%%%%%%%%%%%%%%%%%%%%%%%%%%%%%%%%%%%%%%%%%%%%%%%%%%

To deal with kinetic energy density contributions, we can proceed in analogy to 
Sec.~\ref{SSec:LSDA} and define a generalized kinetic energy density:
\begin{eqnarray}
{\bar\tau} = \frac{1}{2}( {\tau} + \bu \cdot \bsigma )
 = \frac{1}{2} \left( \begin{array}{cc} 
 \tau +  u_z & u_x - i u_y \\
 u_x + i u_y  & \tau - u_z \end{array} \right)  \,,
\label{Eq:GTAU}
\end{eqnarray}
where $\tau$ and $\bu$ can be written in terms of two-component 
spinors as
\begin{eqnarray}
\tau({\bf r}) =  \frac{1}{2} \sum_{i \in occ} \Big(  \bnabla  \Psi_i  ({\bf r}) \Big) ^\dagger \cdot
\bnabla \Psi_i ({\bf r})\,,
\end{eqnarray}
and
\begin{eqnarray}
u_k({\bf r}) = \frac{1}{2} \sum_{i \in occ} \Big(  \bnabla  \Psi_i  ({\bf r}) \Big)  ^\dagger \sigma_k \cdot
\bnabla \Psi_i ({\bf r})\,\,\,\,\,\,\,\,\,\,(k=x,y,z).
\end{eqnarray}
Comparing $\bar\tau$ (Eq.~\ref{Eq:GTAU}) and $\bar n$ (Eq.~\ref{Eq:GN}) one is tempted to define 
$ \tau{\sml +}$ and $ \tau{\sml -}$ as the local eigenvalues of  ${\bar\tau}$. However, this choice
would lead to a different local reference frame than the one used for $\bar n$, and therefore 
collinear solutions obtained in this way will not necessarily be the same 
as those obtained with standard unrestricted KS calculations.
To avoid this problem, we have chosen the following definition for $ \tau{\sml +}$ and $ \tau{\sml -}$,
\begin{eqnarray}
 \tau_{\sml \pm} = \frac{1}{2}(\tau \pm {\hat{\bf m} \cdot \bu }) \,,
\end{eqnarray}
which is equivalent to locally projecting $\bu$ onto the axis defined by $\bf m$.
Using this choice for  $ \tau{\sml +}$ and $ \tau{\sml -}$, 
the contribution to the XC potential matrix elements can be written as
\begin{eqnarray}
(V_{\sml{xc}}^{\sml{NC}})_{\mu\nu}& = & \sum_{j=x,y,z} \int \frac{\partial f}{\partial \tau} (\nabla_j\phi_\mu  
\nabla_j \phi_\nu) d^3r  \nonumber \\
& & +  \sum_{j,k=x,y,z}\int  \frac{\partial f}{\partial u_k} \sigma_k  (\nabla_j\phi_\mu \nabla_j \phi_\nu ) d^3r \nonumber \\
& & + \sum_{k=x,y,z} \int \frac{\partial f}{\partial m_k} \sigma_k  (\phi_\mu \phi_\nu)  d^3r   \,.
\label{Eq:VTAU}
\end{eqnarray}
The partial derivatives of $f$ in Eq.~\ref{Eq:VTAU} can be expressed in terms of 
the derivatives of $f$ with respect to $\tau{\sml +}$ and $\tau{\sml -}$  as
\begin{eqnarray}
\frac{\partial f}{\partial \tau}  = \frac{1}{2} h^{\sml +}\,,
\end{eqnarray}
\begin{eqnarray}
\frac{\partial f}{\partial u_k}  = \frac{m_k}{2 m} h^{\sml -}\,,
\end{eqnarray}
and
\begin{eqnarray}
\frac{\partial f}{\partial m_k}  = 
\frac{1}{2 m } h^{\sml -} 
\sum_{j=x,y,z} u_j ( \delta_{jk} -  \frac{m_j m_k}{m^2} ) \,,
\label{Eq:TAU3}
\end{eqnarray}
where
\begin{eqnarray}
h^\pm=\frac{\partial f}{ \partial \tau_{\sml{+}}} \pm 
\frac{\partial f}{ \partial \tau_{\sml{-}}} \,.
\end{eqnarray}
The first term on the right-hand side of Eq.~\ref{Eq:VTAU} contributes to
the scalar potential $ {\cal{\bm E}}_{xc}^{\sml NC} $, while the second and third terms are spin dependent and therefore
contribute to  $ {\cal{\bm B}}_{xc}^{\sml NC} $. 
In the limit 
where $m\rightarrow0$ then $|{\bf m}\cdot \bu|\rightarrow0$, and therefore  
${\cal{\bm B}}_{xc}^{\sml NC}\rightarrow {\bf 0}$ since $h^-\rightarrow0$.

%%%%%%%%%%%%%%%%%%%%%%%%%%%%%%%%%%%%%%%%%%%%%%%%%%%%%%%%%%%%%%%%%%%%%%%%%%%%%%%%%%%%%%%%%%%%%%
\subsection{Laplacian of the density}
\label{SSec:LAP}
%%%%%%%%%%%%%%%%%%%%%%%%%%%%%%%%%%%%%%%%%%%%%%%%%%%%%%%%%%%%%%%%%%%%%%%%%%%%%%%%%%%%%%%%%%%%%%

The dependence of $E_{xc}^{\sml DFA}$ with the  Laplacian  of the density can be generalized
for the noncollinear case through $\nabla^2 n_{\pm}$, which in terms of $n$, $\bf m$, and their derivatives can
be written as: 
\begin{eqnarray}
\nabla^2 n_{\pm}&  =  & \frac{1}{2}
\Big\{ 
\nabla^2 n \pm \sum_{k=x,y,z} 
\big( 
 \frac{m_k}{m} \nabla^2 m_k  
+  
\sum_{j=x,y,z} ( \frac{(\nabla_j m_k)^2}{m} \nonumber  \\ 
& & -  
\sum_{l=x,y,z}\frac{m_l m_k \nabla_j m_l \nabla_j m_k }{m^3} ) 
\big)
\Big\}  \,.
\label{Eq:LAP}
\end{eqnarray}

Three types of contributions to the XC potential arise in this case, since $\nabla^2 n_{+} $ and
$\nabla^2 n_{-} $  depend on $\nabla^2 n$, $\nabla^2 m_k$, 
$m_k$, and $\bnabla m_k$ (Eq.~\ref{Eq:LAP}).
\begin{eqnarray}
(V_{\sml{xc}}^{\sml{NC}})_{\mu\nu} & = & \int (t^{\sml +} + t^{\sml -} \bsigma \cdot {\hat {\bf m} }  )  \nabla^2 (\phi_\mu \phi_\nu) d^3r  \nonumber \\ 
&  & + \sum_{j=x,y,z} \int  (t^{\sml +} + t^{\sml -} \bsigma \cdot {\hat {\bf m} }  ) (\nabla_j \phi_\mu \nabla_j \phi_\nu) d^3r  \nonumber \\
& & + \sum_{k,j=x,y,z}  \int
   \frac{t^{\sml -} \sigma_k }{2} 
   \frac{\partial (\nabla^2 m)  }{\partial (\nabla_j m_k) } (\nabla_j \phi_\mu \nabla_j \phi_\nu) d^3r \nonumber \\
& & + \sum_{k=x,y,z} \int \frac{t^{\sml -} \sigma_k }{2} 
\frac{\partial (\nabla^2 m ) }{\partial m_k} (\phi_\mu  \phi_\nu)  d^3r
\label{Eq:VLAP}
\end{eqnarray}
where 
\begin{eqnarray}
t^\pm=\frac{\partial f}{ \partial (\nabla^2 n {\sml{+}} ) }  \pm 
\frac{\partial f}{ \partial (\nabla^2 n {\sml{-}} ) } \,.
\end{eqnarray}
The derivatives of $ \nabla^2 m $ with respect to the  linear variables can be obtained from 
Eq.~\ref{Eq:LAP}. 
In the limit of no magnetization where $m\rightarrow0$ and $\nabla^2 m\rightarrow0$,
all contributions to 
${\cal{\bm B}}_{xc}^{\sml NC}$ are zero since $t^-\rightarrow0$.

%%%%%%%%%%%%%%%%%%%%%%%%%%%%%%%%%%%%%%%%%%%%%%%%%%%%%%%%%%%%%%%%%%%%%%%%%%%%%%%%%%%%%%%%%%%%%%
\subsection{Hartree-Fock type exchange}
\label{SSec:HFX}
%%%%%%%%%%%%%%%%%%%%%%%%%%%%%%%%%%%%%%%%%%%%%%%%%%%%%%%%%%%%%%%%%%%%%%%%%%%%%%%%%%%%%%%%%%%%%%

The HF type exchange contribution to $(V_{xc})_{\mu\nu}$, needed for hybrid density functional calculations,
can be generalized in terms of
two-component spinors as 
\begin{equation}
K_{\mu\nu}^{\sigma\sigma'}=\sum_{\lambda\xi}P_{\lambda\xi}^{\sigma\sigma'}
(\mu\lambda|\xi\nu)
\,, \label{K}
\end{equation}
where $P_{\lambda\xi}^{\sigma\sigma'}$ are the spin blocks of the generalized density matrix in Eq.~(\ref{Eq:GP}).
The notation $(\mu\nu|\xi\lambda)$ has been introduced for the two
electron integrals in the AO basis set. This is analogous to
the generalized unrestricted HF (GUHF) approximation.\cite{McWeeny}
The evaluation of the four matrix blocks of $K_{\mu\nu}^{\sigma\sigma'}$
is carried out by splitting $P_{\lambda\xi}^{\sigma\sigma'}$ into real and imaginary
parts, and symmetric and
antisymmetric components. The symmetric imaginary and antisymmetric real
contributions
to $K^{\alpha\alpha}$ and $K^{\beta\beta}$ are zero because of the
hermiticity requirement of the Kohn-Sham (or Hartree-Fock) Hamiltonian. Therefore, 
a total of eight 
HF exchange blocks need to be computed, four of them are symmetric and
four are antisymmetric.

%%%%%%%%%%%%%%%%%%%%%%%%%%%%%%%%%%%%%%%%%%%%%%%%%%%%%%%%%%%%%%%%%%%%%%%%%%%%%%%%%%%%%%%%%%%%%%
\section{Implementation}
%%%%%%%%%%%%%%%%%%%%%%%%%%%%%%%%%%%%%%%%%%%%%%%%%%%%%%%%%%%%%%%%%%%%%%%%%%%%%%%%%%%%%%%%%%%%%%

We have implemented the SCF solution of the GKS equations in the
$Gaussian$ suite of programs.\cite{gdv-e5} Molecular spinors (Eq.~\ref{Eq:Spinors})
are spanned in terms of atomic Gaussian orbitals using  a set of complex 
coefficients $c_{\mu i }^{\sigma}$. These coefficients are employed to 
construct the generalized density matrix $P_{\mu\nu}$    (Eq.~\ref{Eq:GP}), from 
which the Hartree-Fock type exchange matrix can be calculated, as well as all the variables 
needed for the numerical quadrature  employed in the evaluation of  $ (V_{xc}^{\sml NC})_{\mu\nu} $.
To accelerate the SCF convergence,
we have generalized  
the direct inversion of the iterative subspace (DIIS)\cite{DIIS} and the 
energy-based DIIS\cite{EDIIS} techniques  for two-component complex spinors.

In a GKS calculation, the spin density of the system is fully unconstrained, and is thus allowed to
change in any arbitrary spatial direction. 
For instance, in a simple calculation of the (nonrelativistic) ground state of the hydrogen atom using LSDA, one 
can obtain an infinite manifold of solutions with the same total energy but different orientation of 
the spin density. 
These solutions are just linear combinations of the two degenerate linearly
independent solutions.
%This could lead to some unphysical interpretation since the true ground state is 
%two-fold degenerate due to spin. However, one should interpret this infinite manifold of solutions as 
%originated in the arbitrary choice of the spin quantization axis and not as different physical solutions. 
%In other words, in the GKS approach any linear combination of  the two  
%spin-degenerate $\alpha$ and $\beta$ solutions is allowed.

We have verified for a representative sample of 
functionals that for cases with  ground state collinear solutions,  
the GKS solution for different choices
of the quantization axis  gives the same total energy as in a collinear  UKS calculation.
We have also verified that the calculated electric dipole moment evaluated  as
finite differences agrees with the expectation value of the electric dipole operator, satisfying
the Hellmann-Feynman theorem.

% The extra degrees of freedom of the  spin variables also leads to additional convergence problems, 
% and the initial guess employed in 
% the SCF procedure now plays an even more important role than in standard unrestricted calculations. 
% We have found that for the test cases studied in this paper, it is useful to start the GKS SCF iterations 
% from an initial guess generated as a result of a GKS calculation of the same systems in the presence of 
% Fermi contact (FC) type  operators conveniently placed at the nuclear sites ${\bf R}_N$,
% \begin{eqnarray}
% H_{\sml{FC}}=\sum_N \sum_{i=x,y,z} A_i^{\sml N}\,\delta({{\bf r}-{\bf R}_N}) s_i \,,
% \end{eqnarray}
% were $A_i^{\sml N}$ are the amplitudes of the  FC interactions and $s_i$ are the components of the 
% electronic spin operator.
% In this way, the initial density is
% spin polarized in the atomic regions in a similar manner than the target density. This, of course, assumes
% {\em a priori} knowing the final state, and that may not be always the case.

The third term on the right-hand side of Eq.~\ref{Eq:VGGA} leads to instabilities in 
the numerical integration due to the presence of $\bnabla (m_i/m)$,  which is exactly zero for collinear spin densities but 
may present significant  oscillations for spin densities that are slightly noncollinear. 
For some cases, this prevented  converging  the total SCF energy better than 10$^{-6}$ hartree, which is not enough for 
our standard accurate convergence criteria.
To avoid this problem, we 
discard contributions to Eq.~\ref{Eq:VGGA} from grid points where the 
magnetization helicity, defined as $m_h={{\bf m} \cdot (\bnabla \times {\bf m})}$, is less than a certain 
threshold ($m_h=0$ in collinear cases). In our tests, a cutoff value of $m_h<$10$^{-6}$ worked reasonably well.

%%%%%%%%%%%%%%%%%%%%%%%%%%%%%%%%%%%%%%%%%%%%%%%%%%%%%%%%%%%%%%%%%%%%%%%%%%%%%%%%%%%%%%%%%%%%%%
\section{Results}
%%%%%%%%%%%%%%%%%%%%%%%%%%%%%%%%%%%%%%%%%%%%%%%%%%%%%%%%%%%%%%%%%%%%%%%%%%%%%%%%%%%%%%%%%%%%%%

\begin{figure}
   \includegraphics[width=8cm]{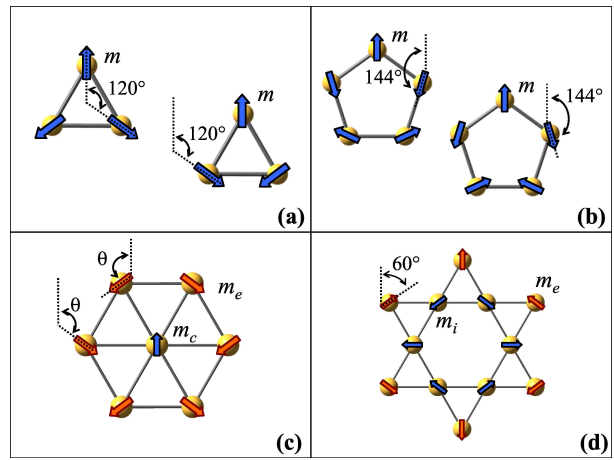}
\caption{ Schematic representation of the Cr$_3$, Cr$_5$, Cr$_7$, and Cr$_{12}$ clusters employed in our tests.
   The arrows represent the magnetization orientation on each atom as qualitatively  obtained in our calculations. 
 For Cr$_3$ and Cr$_5$, two different chiralities were considered.   }
  \label{Fig:Schemes}
\end{figure}

In order
to test our GKS code, we have chosen a set of planar Cr clusters 
where the ground state is expected to exhibit noncollinear spin density 
arising from geometrically  frustrated antiferromagnetic 
coupling (as in a Heisenberg spin Hamiltonian model) between neighboring  Cr atoms. 
% These clusters can be thought as simple model systems where
% noncollinear magnetism takes place.
In Fig.~\ref{Fig:Schemes}, we show a scheme of the Cr$_3$ ($C_{3v}$), Cr$_5$ ($C_{5v}$) , Cr$_7$ ($C_{6v}$), 
and Cr$_{12}$ ($C_{6v}$) clusters and their resulting magnetic structures obtained in this work. 
In all cases, we have set the Cr--Cr bond length 
to 3.70~Bohr in our calculations. This allows us to compare the direct effect of each functional on the magnetization.

%------------------------------------------------------------------------------------------
%------------------------------------------------------------------------------------------

All calculations were carried out using a Ne core energy-consistent 
relativistic effective core potential (RECP)  from the Stuttgart/Cologne group (Ref.~\onlinecite{Stuttgart}).
We have employed a polarized triple-$\zeta$ Gaussian basis set consisting of 8$s$7$p$6$d$1$f$ functions 
contracted to 6$s$5$p$3$d$1$f$.\cite{Stuttgart} 
Even though our implementation offers the possibility of including the spin-orbit operator, either using RECPs or 
in an all-electron framework, we have chosen not to include the spin-orbit interaction in the present
test calculations. Atomic magnetic moments are calculated according to Mulliken population analysis.
The expectation value $\langle S^2\rangle$ is evaluated in all DFT cases as for the GUHF determinant.

%------------------------------------------------------------------------------------------
\begin{table}
    \caption{Atomic magnetic moments (in Bohr magnetons, $\mu_B$) and $\langle S^2\rangle $ (in $\mu_B^2$) of 
    Cr clusters calculated using different
    energy functionals. See Fig.~\ref{Fig:Schemes} for a  scheme of the spin density
    configurations.
    \label{Table:Cr3-5}
    }
\begin{ruledtabular}
  \begin{tabular}{llrrrrr}
  &   &   \multicolumn{5}{c}{Method} \\
\cline{3-7}
\raisebox{1.5ex}[0pt]{Cluster} & \raisebox{1.5ex}[0pt]{Property}
                              &    LSDA       &   PBE   &   TPSS     &  PBEh     &   GUHF   \\
\cline{1-7}
Cr$_3$ ($C_{3v}$)     & $m$   &     1.44      &  1.66   &   1.93     &   2.40    &  2.95   \\   
      & $\langle S^2\rangle$  &    3.25       &   3.87  &   4.71     &   6.32    &  8.11   \\
\cline{1-7}
Cr$_5$ ($C_{5v}$)     & $m$   &     1.61      & 1.84    &   2.07     & 2.47      &  2.91   \\
    &    $\langle S^2\rangle$ &    5.21       &  6.21   &  7.38      &  9.78     &  12.47  \\
\cline{1-7}
Cr$_7$ ($C_{6v}$)     & $m_c$ &    0.18       &   0.21  &  0.31      &  2.09     &  2.36     \\
    &    $m_e$                &  0.24         &  0.89   & 1.29       &  2.33     &  2.87    \\
    &    $\theta$ (deg.)       &   143         &  103    &  100       &   105     &  107     \\
    &    $\langle S^2\rangle$ &   2.08        &   4.02  &   5.80     &   14.60   &  22.57   \\
\cline{1-7}
Cr$_{12}$ ($C_{6v}$)     & $m_i$ & 0.72         &  0.84   &  1.03      &  1.86     & 2.15      \\
    &    $m_e$                 &  1.24        &  1.50   & 1.72       & 2.28      &   2.80   \\
    &    $\langle S^2\rangle$  &  5.73        & 7.60    & 9.67       &   17.71   &   22.82  \\
    \end{tabular}
\end{ruledtabular}
\end{table}
%------------------------------------------------------------------------------------------

We have chosen one representative functional from each class of functionals discussed in Section II. 
For the LSDA, we employ LDA (Dirac) exchange and the parametrization of Wosko, Wilk, and Nusair\cite{vosko80} for correlation (SVWN5); 
for the GGA we use the functional
of Perdew, Burke, and Ernzerhof (PBE);\cite{PBE_1} for the meta-GGA we use the functional developed by
Tao, Perdew, Staroverov, and Scuseria (TPSS);\cite{TPSS} and as a representative hybrid functional we use 
PBEh\cite{PBE0a} (PBE hybrid, also refer to as  PBE1PBE\cite{PBE0b} and  PBE0\cite{PBE0c} in the literature). 
For comparison, we also present  results for  the GUHF case.

%------------------------------------------------------------------------------------------
\begin{figure}
  \begin{center}
      \includegraphics[width=8cm]{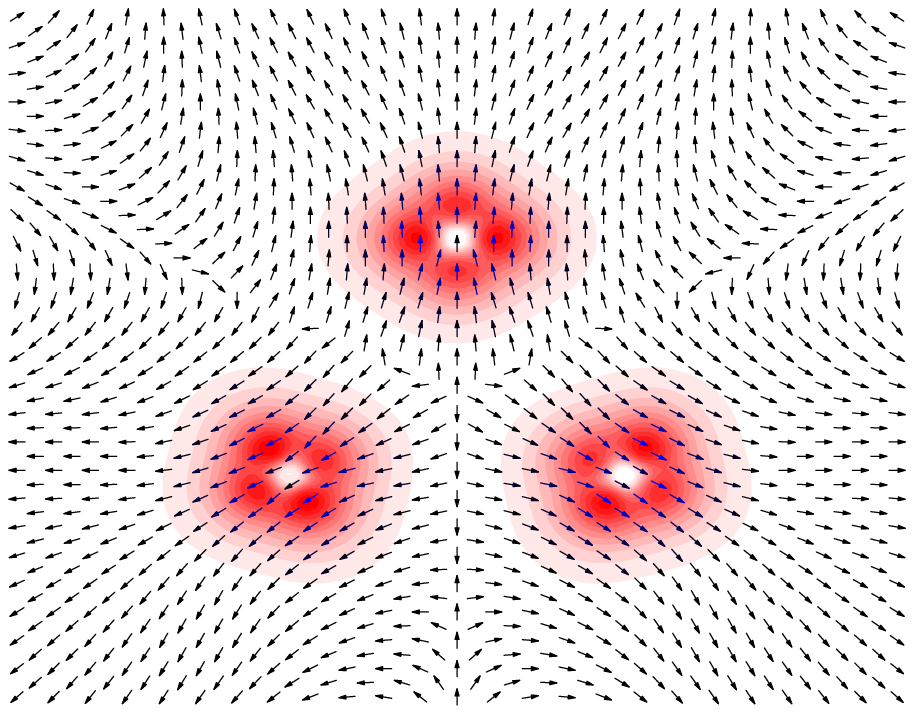}\\
      {\bf (a)}\\
      \includegraphics[width=8cm]{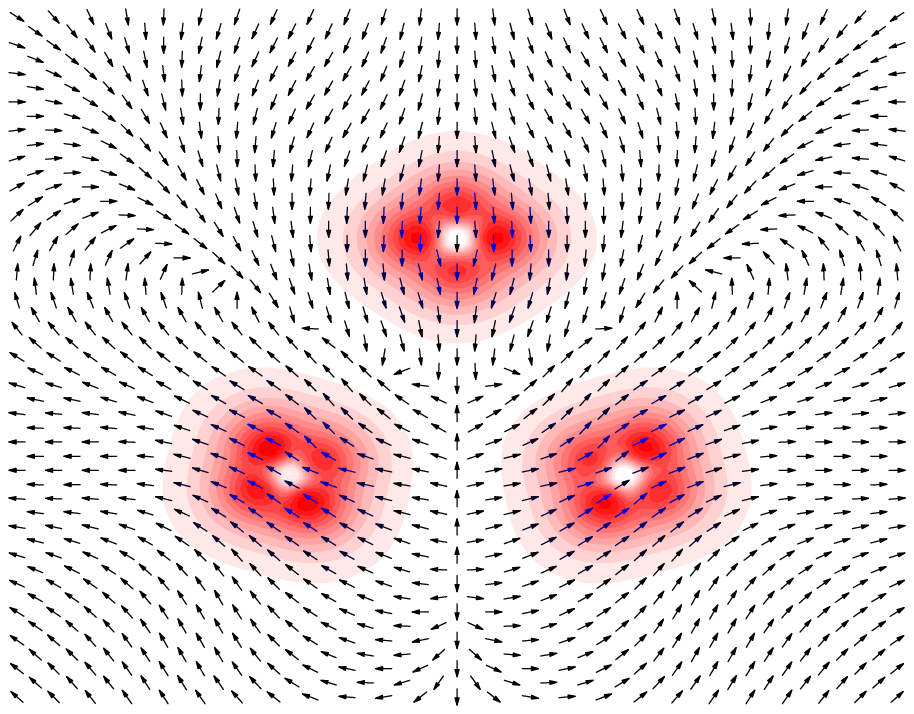} \\
      {\bf (b)}\\
  \end{center}
  \caption{Magnetization plot for Cr$_3$ obtained in  a PBE calculation.
    The arrows show the direction of the spin polarization (${\bf m}/m$) in the plane containing the nuclei, 
    whereas the spin modulus $m$ is represented in red. The top (a)  and bottom (b) panels show two 
    energetically degenerate configurations with + and - chiralities, respectively.  }
  \label{Fig:Cr3}
\end{figure}
%------------------------------------------------------------------------------------------

For Cr$_3$ and Cr$_5$,  we were able to verify that the two chiral magnetic structures (Fig.~\ref{Fig:Schemes}a and 
Fig.~\ref{Fig:Schemes}b, respectively) have the same total energy. These two chiral magnetic states 
can be thought of as a product of a reflection of the spin density pseudovector 
in a molecular symmetry plane. 
For Cr$_3$ and Cr$_5$, we have found that starting the SCF procedure from different initial guesses always leads to coplanar 
spin densities, although the plane containing the spin density does not necessarily coincide with the plane containing the nuclei
since the spin density can arbitrarily rotate without changing the total energy.
%(the situation would be different if spin-orbit is included in the calculation because 
% it would ``lock'' spatial and spin degrees of freedom). 
We therefore have chosen to constrain the spin magnetization to  the
plane containing the nuclei for the rest of our tests. 

\begin{figure}
  \begin{center}
      \includegraphics[width=8cm]{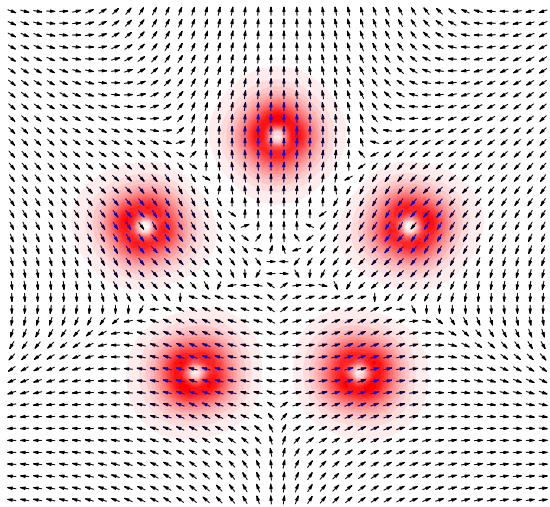}\\
      {\bf (a)}\\
      \includegraphics[width=8cm]{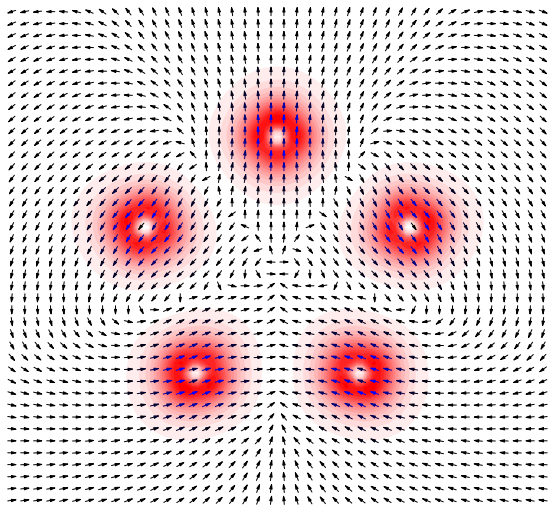} \\
      {\bf (b)}\\
  \end{center}
  \caption{Magnetization plot for Cr$_5$ obtained in a PBE calculation.
    The arrows show the direction of the spin polarization (${\bf m}/m$) in the plane containing the nuclei,
    whereas the spin modulus $m$ is represented in red. The top (a)  and bottom (b) panels show two
    energetically degenerate configurations with + and - chiralities, respectively. }
  \label{Fig:Cr5}
\end{figure}

In Figs.~\ref{Fig:Cr3} and \ref{Fig:Cr5}, we present a plot of the PBE spin density
in the plane containing the nuclei for Cr$_3$ and Cr$_5$, respectively. 
The white (low spin polarization) holes at the nuclear positions are a consequence 
of the pseudopotential approximation. The red  zones surrounding the nuclei
correspond to high spin polarization regions. Four lobes can be distinguished around each 
atomic center, which is a signature of the spin polarization of the $d$ orbitals.

%------------------------------------------------------------------------------------------
% ------------------------------------------------------------------------------------------

From Figs.~\ref{Fig:Cr3} and \ref{Fig:Cr5},
it can also be seen that the magnetization tends to be collinear in the atomic regions.
Inside these atomic domains, the magnetization angle changes smoothly  whereas it
changes abruptly at the domain boundary.
This was also observed in Fe clusters\cite{Oda_1998}  
and in unsupported  Cr monolayers in the 120$^\circ$  N\'eel state.\cite{Kurz_2004}
Spin density plots obtained using density functionals other than PBE do not differ qualitatively from
these plots.  However, the magnitude of the spin polarization does, as it is discussed  below.

In Table~\ref{Table:Cr3-5},  we summarize  the results obtained for Cr clusters 
with the different functionals. In all cases, the atomic magnetization  
increases systematically when going from 
LSDA$\rightarrow$PBE$\rightarrow$TPSS$\rightarrow$PBEh$\rightarrow$GUHF.
The value of $\langle S^2\rangle$ follows the same trend, and it can be taken as a
measure of the total magnetization.
As a remark, we would like to recall that the Cr cluster geometries are fixed in these test calculations and
hence  relaxation effects are not included in the reported atomic magnetic moments.  
 For Cr$_3$ and Cr$_5$,  we obtain 
comparable values of the  atomic magnetization $m$ for a given functional. 
This is not the case for 
Cr$_7$ and Cr$_{12}$ clusters, where the atomic magnetic moments of the internal Cr atoms are  smaller than
those of the external atoms in all cases. From the values of $m_c$ and $m_e$ (see Fig.~\ref{Fig:Schemes}) 
for Cr$_7$ with PBEh and GUHF, we can 
notice a large effect of Hartree-Fock  exchange compared to the rest of the clusters. In fact, GUHF 
gives an atomic magnetization in Cr$_7$ approximately 10 times larger than those obtained with  LSDA,
while for Cr$_3$ and Cr$_5$ the ratio is less than 2, and for Cr$_{12}$ is about 3.

%------------------------------------------------------------------------------------------
\begin{figure}
  \begin{center}
      \includegraphics[width=8cm]{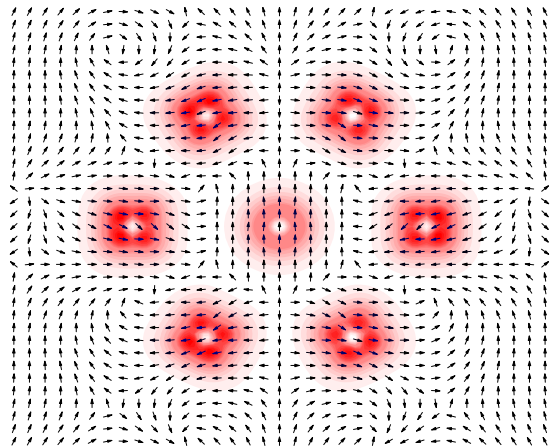}\\
  \end{center}
  \caption{Magnetization plot for the Cr$_7$ cluster obtained in a PBE calculation.
    The arrows show the direction of the spin polarization (${\bf m}/m$) in the plane containing the nuclei,
    whereas the spin modulus $m$ is represented in red. }
  \label{Fig:Cr7}
\end{figure}
%------------------------------------------------------------------------------------------

\begin{figure}
  \begin{center}
      \includegraphics[width=8cm]{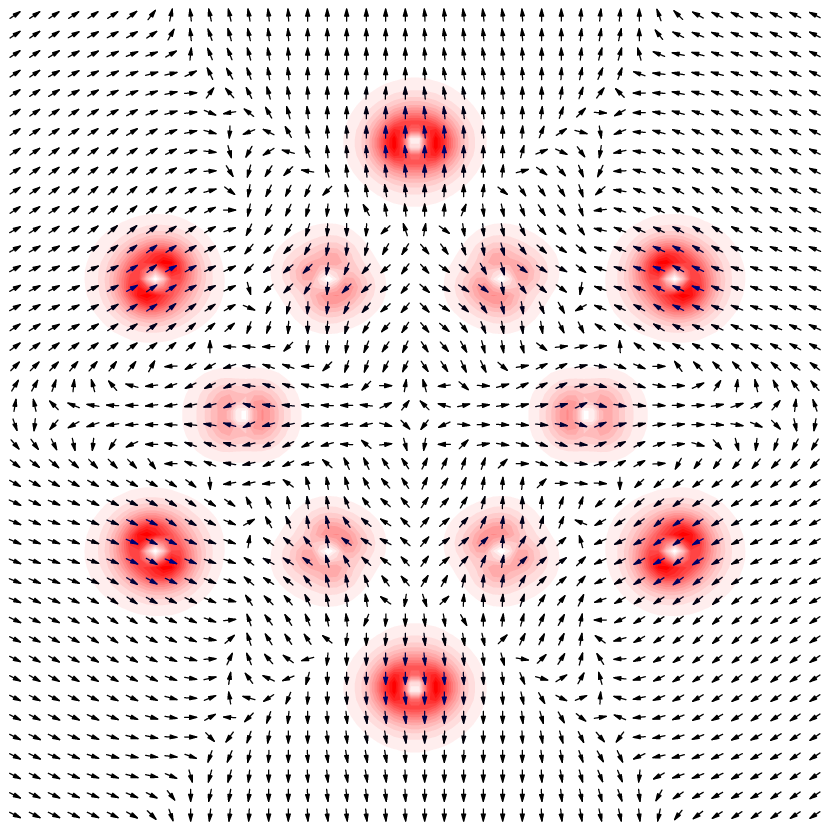}\\
  \end{center}
  \caption{Magnetization plot for the Cr$_{12}$ cluster obtained ina PBE calculation.
    The arrows show the direction of the spin polarization (${\bf m}/m$) in the plane containing the nuclei,
    whereas the spin modulus $m$ is represented in red.}
  \label{Fig:Cr12}
\end{figure}

In Figs.~\ref{Fig:Cr7} and \ref{Fig:Cr12}, we present  the PBE spin density
in the plane containing the nuclei for Cr$_7$ and Cr$_{12}$, respectively.
The plot for  Cr$_7$  resembles the spin density 
in the usupported  Cr monolayer shown in Ref.~\onlinecite{Kurz_2004}. The difference arises mainly in the 
low spin polarization region of the Cr$_7$ cluster.  Cr$_{12}$ can be thought as a cluster model of a 
two-dimensional Kagom\'e lattice. However, as the magnetization of the internal and external atoms is
not the same, one would expect that the actual ground state is a result of different competing effects. 
For instance, in a Heisenberg spin Hamiltonian, 
the uppermost Cr atom, Fig.~\ref{Fig:Schemes}d, couples antiferromagnetically with its weaker polarized 
nearest neighbors and
with its stronger polarized second  nearest neighbors, and it is not clear {\em a priori} which coupling 
is larger. Therefore, we expect that this type of noncollinear DFT calculations would be helpful to 
investigate the magnetic properties of clusters and molecules where a simple Heisenberg spin Hamiltonian cannot be
straightforwardly applied.

%------------------------------------------------------------------------------------------
%------------------------------------------------------------------------------------------

%%%%%%%%%%%%%%%%%%%%%%%%%%%%%%%%%%%%%%%%%%%%%%%%%%%%%%%%%%%%%%%%%%%%%%%%%%%%%%%%%%%%%%%%%%%%%%
\section{Summary and Conclusions}
%%%%%%%%%%%%%%%%%%%%%%%%%%%%%%%%%%%%%%%%%%%%%%%%%%%%%%%%%%%%%%%%%%%%%%%%%%%%%%%%%%%%%%%%%%%%%%

We have generalized the treatment of the electronic spin degrees of freedom in density functional calculations to 
the case where the vector variables employed in the definition of the XC energy can vary in any direction. 
Our noncollinear generalization can be applied 
to general functionals containing a variety of ingredients.
%not 
%only to the LSDA and GGA but also to meta-GGA energy functionals that depend on the kinetic energy density and
%%the Laplacian of the density. We are also able to deal with hybrid density functionals by incorporating
%a portion of generalized Hartree-Fock type exchange in the energy expression in analogy with 
%the generalized unrestricted Hartree-Fock approximation. 
%We have also given a detailed expression for the generalized XC potentials matrix elements.
%
Our generalization assumes that  the XC energy depends on the local variables  in the same manner as in 
the standard collinear case, and that the energy expression is  invariant under rigid rotations of 
the spin quantization axis. This is  not the most general way to define energy functionals for 
noncollinear magnetic systems, but it provides a general starting point to incorporate new terms 
like those suggested in Refs.~\onlinecite{Kleinman_1999} and \onlinecite{capelle}.

% and 
%shown in the derivation 
% that for the GGA and meta-GGA cases the XC magnetic potential is not necessarily parallel to the spin density. 

Test calculations on planar Cr clusters suggest that the choice of energy functional has an important
impact on the resulting atomic magnetic moments, giving qualitatively similar but quantitatively different results. 
We expect that our generalization will open the door to
studies on the performance of density functionals other than LSDA for noncollinear magnetic systems.

%%%%%%%%%%%%%%%%%%%%%%%%%%%%%%%%%%%%%%%%%%%%%%%%%%%%%%%%%%%%%%%%%%%%%%%%%%%%%%%%%%%%%%%%%%%%%%
\section{Acknowledgements}
%%%%%%%%%%%%%%%%%%%%%%%%%%%%%%%%%%%%%%%%%%%%%%%%%%%%%%%%%%%%%%%%%%%%%%%%%%%%%%%%%%%%%%%%%%%%%%

J.E.P thanks K. Capelle, R. Pino, A. Izmailov, and O. Hod for useful discussions
and T. Van Voorhis for suggesting the Cr$_{12}$ example.
This work was supported by the Department of Energy
Grant No. DE-FG02-01ER15232, ARO-MURI DAAD-19-3-1-0169,
and the Welch Foundation.

%--------------------------------------------------------------------------
\bibliography{references}
%--------------------------------------------------------------------------

\end{document}